\documentclass[11pt,a4paper,twocolumn]{article}

\usepackage[utf8]{inputenc}
\usepackage{times}
\usepackage[colorlinks=false, pdfborder={0 0 0}]{hyperref}
\usepackage{breakurl}
\usepackage[rightcaption]{sidecap}
\usepackage{stfloats}

\usepackage{amsmath}
\usepackage{listings}
\usepackage{eurosym}
\usepackage{afterpage}

\usepackage{enumitem}

\usepackage[font=small,margin=1ex,labelfont=bf]{caption}

\usepackage{subfigure}

\usepackage{calc}
\usepackage[left=2.0cm,right=2.0cm,top=3.0cm,bottom=3.0cm]{geometry}

\usepackage{parskip}
\usepackage{graphicx}

\usepackage{fancyhdr}
\fancypagestyle{firstpage}{
  \fancyhead{}
  \cfoot{\footnotesize{\em Technischer Bericht Nr.~2016-01, Hochschule Niederrhein, Fachbereich Elektrotechnik \& Informatik (2016)}}
  
}

\title{\vspace*{-10mm}Estimating Wealth Distribution: Top Tail and Inequality}
\author{
Christoph Dalitz\\
Institut f\"ur Mustererkennung\\
Hochschule Niederrhein\\
Reinarzstr. 49, 47805 Krefeld\\
{\tt christoph.dalitz{@}hsnr.de}
}
\date{}

\begin{document}

\renewcommand{\labelenumi}{\arabic{enumi})}

\twocolumn[
  \begin{@twocolumnfalse}
    \maketitle
\begin{abstract}
This article describes mathematical methods for estimating the top-tail 
of the wealth distribution and therefrom the share of total wealth that the 
richest $p$ percent hold, which is an intuitive measure of inequality. As the 
data base for the top-tail of the wealth distribution is inevitably less 
complete than the data for lower wealth, the top-tail distribution is 
replaced by a parametric model based on a Pareto distribution. The 
different methods for estimating the parameters are compared and new simulations
are presented which favor the maximum-likelihood estimator for the Pareto
parameter $\alpha$. New criteria for the choice of other parameters are
presented which have not yet been discussed in the literature before.
The methods are applied to the 2012 data from the ECB Household and
Consumption Survey (HFCS) for Germany and the corresponding rich list from
the Manager Magazin. In addition to a presentation of all formulas,
{\em R} scripts implementing them are provided by the author.
\end{abstract}
\vspace*{2ex}

  \end{@twocolumnfalse}
  ]

\thispagestyle{firstpage}

\section{Introduction}
For some time, economists have paid little attention to questions of inequality and wealth distribution. The Nobel Prize economist Robert Lucas, e.g., declared \cite{lucas04}: ``Of the tendencies that are harmful to sound economics, the most seductive, and in my opinion, the most poisonous, is to focus on questions of distribution.'' According to Robert H.~Wade \cite{wade14}, this attitude was due to the dominating theory of trickle-down economics, which celebrates inequality as an incentive for effort and creativity. 

The ``publishing sensation'' (Wade) of Piketty's ``Capital in the Twenty-First Century'' \cite{piketty14} has brought back inequality into the focus of mainstream economists. Piketty did not examine why inequality matters (see \cite{wilkinson09} or \cite{stiglitz12} for this topic), but how it evolved over time. Most of his data stemmed from tax records, which go back to the beginning of the 19th century in some cases. While tax data have the advantage that their collection is obligatory, they have other shortcomings. For about thirty years, governments in most OECD countries have followed the consensus among mainstream economists that taxes on the rich and aid to the poor tend to be inversely correlated with economic growth. This has resulted in tax cuts for the rich and the elimination of some types of tax: the wealth tax, e.g., was dropped in Germany in 1997, which has the side effect that some information about wealth can no longer be drawn from tax data. Another problem with tax records is bias at the top tail because higher income and wealth are more likely to be hidden in tax havens.

Alternative data sources are surveys like the Eurosystem Household and Finance Consumption Survey (HFCS) \cite{ecb13}. Although the HFCS combines comprehensive information in a unified form across several European countries, it suffers from bias at the top tail, too. One reason for this bias is the random sampling process which is very unlikely to represent the rare values from the top tail of the distribution. Another reason is that the survey response rate is known to be lower for higher income and wealth \cite{kennickell93}. These effects make it necessary to correct for the missing rich in some way.

To this end, Vermeulen \cite{vermeulen14} proposed to replace the survey data for high wealth values with a parametric model based on the Pareto distribution, i.e., a probability distribution with density $f(w)=Cw^{-(\alpha+1)}$. For parameter estimation, he suggested to combine the survey data with rich lists like the Forbes World's billionaires list. Bach et al.~\cite{bach15} applied this idea to the HFCS data in combination with national rich lists for different European countries. Eckerstorfer et al.~\cite{eckerstorfer15} relied only on the HFCS data, from which they extrapolated the top-tail of the wealth distribution with well-defined criteria for the choice of some model parameters, which have been chosen in an ad-hoc manner by Vermeulen and Bach et al. All authors used these methods to estimate the wealth share of the richest one-percent, which they found to increase with respect to the raw HFCS data when the correction for the missing rich is applied. The actual values for this share were estimated to 33\% for Germany \cite{bach15} and 38\% for Austria \cite{eckerstorfer15}.

This article provides a comprehensive summary of the statistical model and how it is used to estimate the wealth share of the top one percent. It is organized as follows: section \ref{sec:database} describes the underlying HFCS data and the rich list for Germany. Section \ref{sec:model} describes the statistical model and how it is utilized for computing the percentile wealth share. Section \ref{sec:parameter} gives a survey of methods for estimating the model parameters. It should be noted, that criteria for determining some of these parameters have not yet been discussed in the literature. The results of these methods when applied to the data for Germany are presented in section \ref{sec:results}. The {\em R} code written for the present study can be downloaded from the author's website\footnote{\url{http://informatik.hsnr.de/~dalitz/data/wealthshare/}}.

\section{Data base}
\label{sec:database}
This study is based on two data sources: the Eurosystem Household Finance and Consumption Survey (HFCS) 2012, and the Manager Magazin rich list from 2012 with extended information collected by the Deutsches Institut für Wirtschaftsforschung (DIW).

The HFCS was performed between 2008 and 2011 by the national banks in the Eurosystem countries Belgium, Germany, Spain, France, Italy, Greece, Cyprus, Luxembourg, Malta, Netherlands, Austria, Portugal, Slovenia, Slovakia, and Finland. In each country, a questionnaire was sent to sample households on basis of the sampling criteria described in \cite{ecb13}. From all the data collected, the present study only uses the {\em net wealth} (variable DN3001) for country Germany (variable SA0100 = DE). To compensate for sampling errors, two counter-measures were taken by the national banks:
\begin{itemize}
\item As rich households are known to have a lower response rate, rich households were over-sampled on basis of geographical area.
\item To each response, a {\em household weight} (variable HW0010) was assigned that estimates the number of households that this particular household represents. The rough idea, but no details, of the weighting process is described in \cite{ecb13}.
\end{itemize}
There is some controversy, in which situations the weights should be used. Bach et al.~\cite{bach15} heavily relied on the weights because they used linear regression for estimating the Pareto parameter $\alpha$. Eckerstorfer et al.~\cite{eckerstorfer15} however ignored the weights in some calculations to ``limit the influence of [..] unknown implicit assumptions.'' The present study uses weights throughout.

To deal with item non-response, the HFCS data also provides imputed values for missing variable values. This only affects the variable {\em net wealth}, but not the household weights. For each missing value, five imputed values are provided, such that there are five different complete data sets. These can be used in two different ways. One is by averaging the variables for each household over the five sets, a method used by Bach et al.~in \cite{bach15}. This is called the {\em averaged HFCS data} in this paper. The other method is to take each imputed data set separately to eventually obtain a range for the observables of interest, a method used by Eckerstorfer et al.~in \cite{eckerstorfer15}.

The Manager Magazin annually publishes a list of the 500 richest families in Germany. Their net wealth is estimated from different sources, which include information provided by the person themselves. The editors of the list indicate that the list is incomplete because some persons have asked for removal from the list. To make the information based on families compatible with the HFCS data based on households, Bach et al.~have collected information from public sources about the number of households for each family. Moreover they have identified non-residents on the list and recommend to only use the top 200 entries from this list. The present study uses these data as provided by Bach et al., but with the non-residents removed, and down to the wealth threshold of the 200th entry, which is 500 Mio \euro{} and thus goes down to 206 entries.

From the large gap between the highest HFCS reported wealth (76 Mio \euro{}) and the lowest value in the Manager Magazin list, it can be concluded that oversampling and weights cannot fully compensate for top-tail bias, and that a compensation for the missing rich is necessary.

\section{Statistical model}
\label{sec:model}
The wealth distribution can be represented by a probability density $f(w)$, such that $f(w)\,dw$ is the fraction of households having wealth in $[w,w+dw)$. When this distribution is known, all interesting observables can be computed therefrom. The {\em mean wealth} $\overline{w}$, e.g., is given by
\begin{equation}
\overline{w} = \int_{-\infty}^\infty w\,f(w)\,dw
\end{equation}
and the {\em total wealth} is $N\cdot\overline{w}$, where $N$ is the total number of households. The {\em percentile} $w_p$, i.e.~the wealth value for which $p$ percent of the households are richer, is implicitly defined by
\begin{equation}
\label{eq:w_p}
p = \int_{w_p}^\infty f(w)\,dw
\end{equation}
and from this value, the wealth share $s_p$ of the richest $p$ percent are given by
\begin{equation}
\label{eq:toppart}
s_p = \frac{\int_{w_p}^\infty w\,f(w)\,dw}{\int_{-\infty}^\infty w\,f(w)\,dw}
\end{equation}

\subsection{Density estimation}
The density $f(w)$ must be estimated from the HFCS data or the rich list, which both provide lists of wealth values $w_1<w_2<\ldots<w_k$ and corresponding weights $n(w_i)$, such that $w_i$ is the wealth of the $i$-th sample household and $n(w_i)$ is the number of households that this value represents. The approach in \cite{vermeulen14,bach15,eckerstorfer15} is to approximate $f(w)$ below some threshold $w_0$ with a histogram having cells centered at $(w_i)_{i=1}^k$ and to use a parametric model based on the Pareto distribution for $w>w_0$. This results in the following histogram density estimator for $w \leq w_0$:
\begin{equation}
\label{eq:f_hist}
\hat{f}(w) = \frac{2\cdot n(w_i)}{N\cdot (w_{i+1}-w_{i-1})}
\end{equation}
for each histogram cell centered around $w_i$, i.e., $(w_{i-1}+w_i)/2 < w \leq (w_i + w_{i+1})/2$. Here, $N$ is the total sum over all weights, i.e., $N=\sum_i n(w_i)$. For $w>w_0$, the density is approximated with a Pareto distribution:
\begin{equation}
\label{eq:f_pareto}
\hat{f}(w) = C\cdot w^{-(\alpha+1)}
\end{equation}

\begin{figure}[t]
\includegraphics[width=\columnwidth]{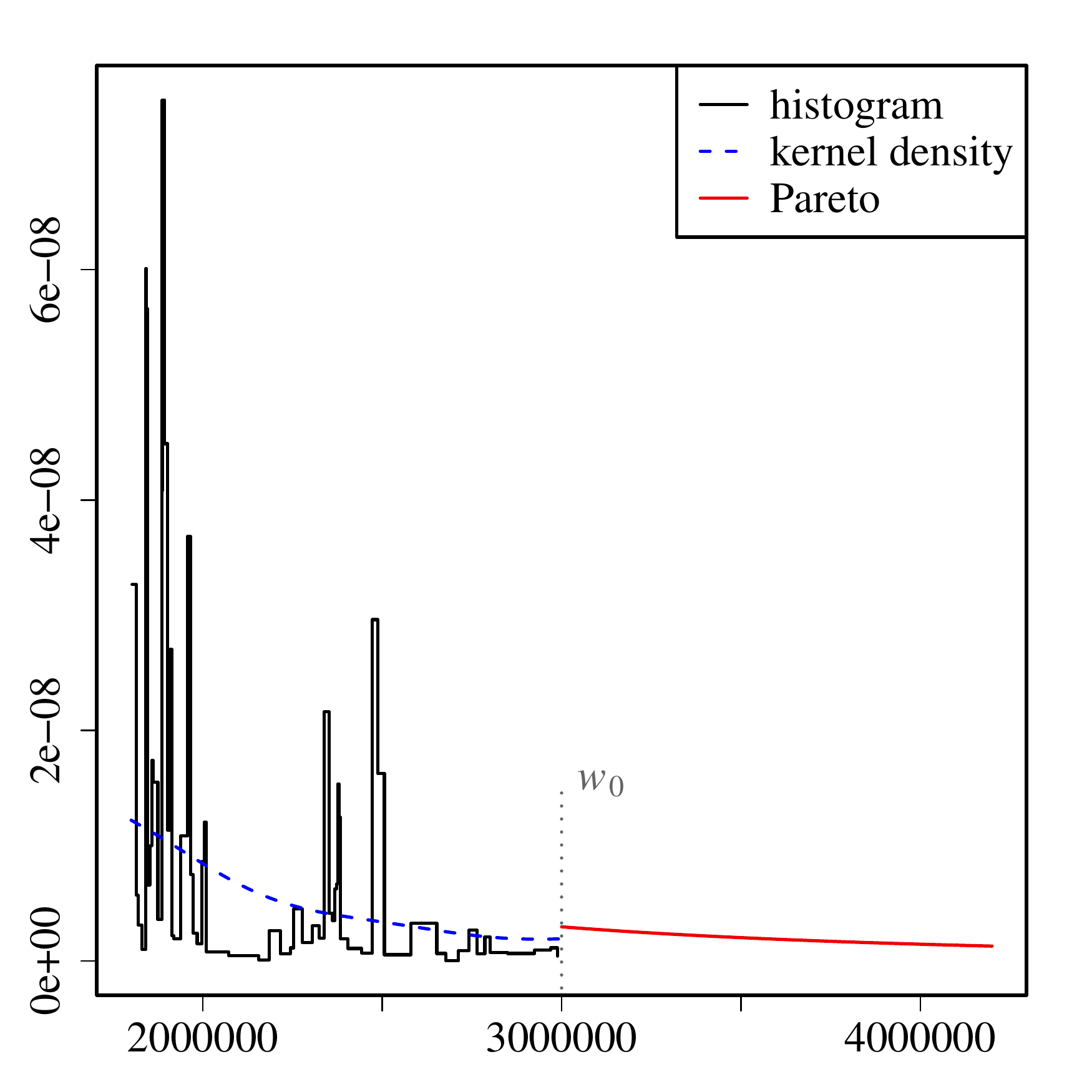}
\caption{\label{fig:densityapprox}Density approximation with Eqs.~(\ref{eq:f_hist}) \& (\ref{eq:f_pareto}) from the averaged HFCS data for Germany with $w_0=3\cdot 10^6$ \euro{}. For comparison, the kernel density according to Eq.~(\ref{eq:f_kernel}) is shown with bandwidth $h=0.25\cdot 10^6$ \euro{}.}
\end{figure}

As can be seen in Fig.~\ref{fig:densityapprox}, the histogram obtained from the HFCS data is very noisy in the transition region. A smoother estimate for $f(x)$ can be obtained with a weighted kernel density estimator \cite{gisbert03}
\begin{equation}
\label{eq:f_kernel}
\hat{f}_{\mbox{\scriptsize\it kern}}(w) = \frac{1}{Nh}\sum_i n(w_i)\cdot K\!\left(\frac{w-w_i}{h}\right)
\end{equation}
where $K(x)=(2\pi)^{-1/2} e^{-x^2/2}$ and $h$ is a band-width. This approximation, however, introduces yet another parameter $h$ and makes a numeric integration for computing the integrals in Eq.~(\ref{eq:toppart}) necessary.

The model given by Eqs.~(\ref{eq:f_hist}) \& (\ref{eq:f_pareto}) has three parameters:
\begin{itemize}\itemsep-0.5ex
\item the threshold $w_0$
\item the power $\alpha$ in the Pareto distribution
\item the normalization constant $C$
\end{itemize}
Methods for determining these parameters are presented in section \ref{sec:parameter}.

\subsection{Percentile share computation}
When the model (\ref{eq:f_hist}) \& (\ref{eq:f_pareto}) is inserted into Eq.~(\ref{eq:w_p}), the percentile $w_p$ is given by (again $N=\sum_i n(w_i)$)
\begin{equation}
\label{eq:w_p_a}
1-p = \frac{1}{N}\sum_{w_i < w_p} n(w_i)
\end{equation}
if the resulting $w_p$ is less than $w_0$. For $w_p>w_0$, it is given by
\begin{equation}
p = \int_{w_p}^\infty Cw^{-(\alpha+1)}\,dw = \frac{C}{\alpha}\cdot w_p^{-\alpha}
\end{equation} 
The resulting wealth share (\ref{eq:toppart}) is for $w_p>w_0$:
\begin{equation}
\label{eq:s_p_a}
s_p= \frac{\frac{C}{\alpha-1}\, w_p^{-(\alpha-1)}}{\sum\limits_{w_i< w_0} \frac{n(w_i)}{N} \;+\; \frac{C}{\alpha-1}\, w_0^{-(\alpha-1)}}
\end{equation}
and for $w_p<w_0$:
\begin{equation}
\label{eq:s_p_b}
s_p = \frac{\sum\limits_{w_p\leq w_i< w_0}\frac{n(w_i)}{N} \;+\; \frac{C}{\alpha-1}\, w_0^{-(\alpha-1)}}{\sum\limits_{w_i< w_0} \frac{n(w_i)}{N} \;+\; \frac{C}{\alpha-1}\, w_0^{-(\alpha-1)}}
\end{equation}
It should be noted that the integrals over the Pareto distribution have been evaluated analytically to obtain these formulas. Bach et al.~\cite{bach15} have instead ``imputed'' wealth values for $w>w_0$ with weights drawn from the Pareto distribution. This leads to the same results, because this imputation is basically a numeric integration.

Eckerstorfer et al.~observed that the upper limit in the integrals in (\ref{eq:toppart}) actually is not infinity, but some finite value $w_{\mbox{\scriptsize\it max}}$. In this case, the term $w_0^{-(\alpha-1)}$ in (\ref{eq:s_p_a}) \& (\ref{eq:s_p_b}) has to be replaced with $(w_0^{-(\alpha-1)} - w_{\mbox{\scriptsize\it max}}^{-(\alpha-1)})$. The upper limit $w_{\mbox{\scriptsize\it max}}$ can be estimated from a rich list as the highest value plus half the distance to the second highest value. For Germany, the Manager Magazin rich list leads to $w_{\mbox{\scriptsize\it max}}\approx 20\cdot 10^9$ \euro{}, so that a truncation at this value only affects the fourth decimal place.

\subsection{Model validation}
While the approximations (\ref{eq:f_hist}) or (\ref{eq:f_kernel}) are non-parametric and make no assumptions about the shape of the wealth density $f(w)$, the approximation (\ref{eq:f_pareto}) only makes sense when the density is actually close to the density of the Pareto distribution for $w>w_0$. This assumption can be verified by testing whether the average wealth above a threshold $w_0$ follows {\em van der Wijk's law}:
\begin{equation}
\label{eq:vanderwijk}
\overline{W}_{\lfloor w_0\rfloor} = \frac{\int_{w_0}^\infty w\, f(w)\, dw}{\int_{w_0}^\infty f(w)\, dw} = \frac{\alpha}{\alpha-1}\cdot w_0
\end{equation}
While it is easy to see that van der Wijk's law holds for the Pareto density by inserting $f(w)=Cw^{-(\alpha+1)}$ on the left hand side of Eq.~(\ref{eq:vanderwijk}), it is also the other way around: the Pareto distribution is the only distribution for which (\ref{eq:vanderwijk}) holds\footnote{To see this, differentiate (\ref{eq:vanderwijk}) with respect to $w_0$ and solve the resulting differential equation for the complementary cumulative distribution function $\overline{F}(w_0)=\int_{w_0}^\infty f(w)\,dw$.}.

\begin{figure}[t]
\includegraphics[width=0.48\columnwidth]{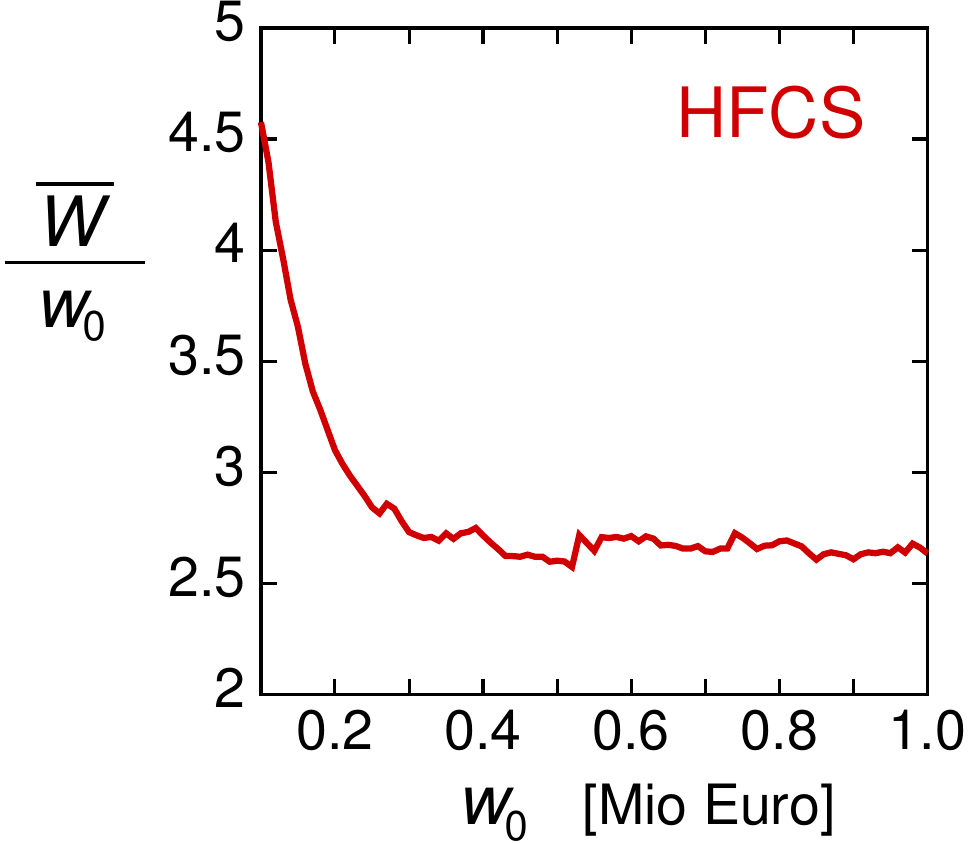}
\includegraphics[width=0.48\columnwidth]{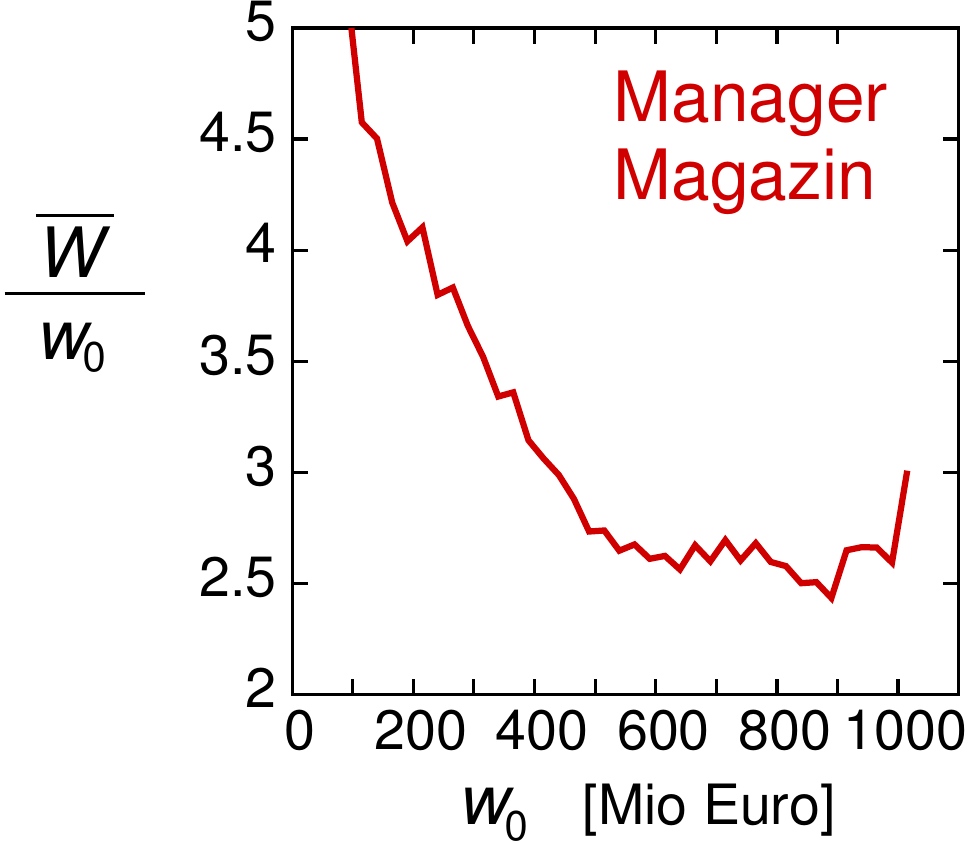}
\caption{\label{fig:wijk}Test of van der Wijk's law for the averaged HFCS data and the Manager Magazin rich list for Germany.}
\end{figure}

Eq.~(\ref{eq:vanderwijk}) thus provides a simple test whether the tail of the survey data follows a Pareto distribution: compute
\begin{equation}
\label{eq:vanderwijk_est}
\frac{\overline{W}_{\lfloor w_0\rfloor}}{w_0} = \frac{\sum_{w_i>w_0}w_i\cdot n(w_i)}{w_0\sum_{w_i>w_0}n(w_i)}
\end{equation}
and check whether it is constant. As can be seen in Fig.~\ref{fig:wijk}, van der Wijk's law holds for the HFCS data for $w_0>0.5\cdot 10^6$ \euro{} and for the Manager Magazin data for $w_0>500\cdot 10^6$ \euro{}. This seems to be a contradiction, because it seems to imply that somewhere above the highest HFCS wealth (76 Mio \euro{}) the distribution deviates from the Pareto shape. It is however in accordance with Bach's observation that the Manager Magazin data tends to become unreliable for more than the 200 highest entries \cite{bach15}, which corresponds to wealths below 500 Mio \euro{}.

\section{Parameter determination}
\label{sec:parameter}
The focus in the published literature so far has been on the determination of the power $\alpha$ in the Pareto distribution, which is summarized in section \ref{sec:parameter:alpha}. The subsequent sections discuss methods for determining the normalization constant $C$ and propose a new method for a unique choice for the transition threshold $w_0$, a problem to which little attention has been given previously.

\subsection{Power $\alpha$}
\label{sec:parameter:alpha}
A simple estimator for $\alpha$ can be obtained by solving van der Wijk's law (\ref{eq:vanderwijk}) \& (\ref{eq:vanderwijk_est}) for $\alpha$:
\begin{equation}
\label{eq:alpha_wijk}
\hat{\alpha}_{\mbox{\scriptsize\it wijk}} = \left[1-\frac{w_{\mbox{\scriptsize\it min}}N(w_{\mbox{\scriptsize\it min}})}{\sum_{w_i\geq w_{\mbox{\scriptsize\it min}}}w_i\cdot n(w_i)}\right]^{-1}
\end{equation}
where $w_{\mbox{\scriptsize\it min}}$ is the wealth threshold, above which the data is used for estimating $\alpha$, and $N(w_{\mbox{\scriptsize\it min}}) = \sum_{w_i\geq w_{\mbox{\scriptsize\it min}}}n(w_i)$ is the sum of all weights greater than $w_{\mbox{\scriptsize\it min}}$. The Monte-Carlo experiments in section \ref{sec:results:simulation} indicate that this estimator has both considerable bias and variance.

An estimator obtained from the maximum-likelihood (ML) principle is therefore preferable, because ML-estimators are generally known to be consistent \cite{clauset09}. The weights $n(w_i)$ can be incorporated into the ML estimator by treating each measured value $w_i$ as if it were measured $n(w_i)$ times. This yields the estimator \cite{vermeulen14}
\begin{equation}
\label{eq:alpha_ml}
\hat{\alpha}_{ml} = \left[\sum_{w_i\geq w_{\mbox{\scriptsize\it min}}} \frac{n(w_i)}{N(w_{\mbox{\scriptsize\it min}})}\ln\left(\frac{w_i}{w_{\mbox{\scriptsize\it min}}}\right)\right]^{-1}
\end{equation}
Another estimator for $\alpha$ can be obtained from the complementary cumulative distribution function of the Pareto distribution:
\begin{displaymath}
\frac{N(w_i)}{N} \approx P(W>w) = \int_w^\infty f(w)\, dw = \frac{C}{\alpha}\cdot w^{-\alpha}
\end{displaymath}
\begin{equation}
\label{eq:pareto_ccdf}
\Rightarrow \quad\frac{N(w_i)}{N(w_{\mbox{\scriptsize\it min}})} \approx \left(\frac{w_i}{w_{\mbox{\scriptsize\it min}}}\right)^{-\alpha}
\end{equation}
Taking the logarithm on both sides leads to a linear relationship $y_i=-\alpha x_i$, from which $\alpha$ can be estimated with a least squares fit as the linear regression coefficient
\begin{equation}
\label{eq:alpha_reg}
\hat{\alpha}_{reg} = \sum_i x_i y_i \left/ \sum_i x_i^2\right. \quad\mbox{ with}
\end{equation}
\begin{displaymath}
y_i = \ln\left( \frac{N(w_i)}{N(w_{\mbox{\scriptsize\it min}})}\right)
\mbox{ and }\; x_i = \ln\left(\frac{w_i}{w_{\mbox{\scriptsize\it min}}}\right)
\end{displaymath}
When the linear regression is not done with Eq.~(\ref{eq:alpha_reg}), but with built in routines of a statistical software package, it is important to set the ``intercept term'' to zero, because otherwise the least square fit is done with a different formula that additionally estimates a constant term\footnote{In {\em R}, this is achieved by using the formula ``$y \sim 0 + x$'' in the function {\em lm}.}.

Clauset et al.~\cite{clauset09} made Monte-Carlo experiments to compare the least squares fit estimator $\hat{\alpha}_{reg}$ with the maximum likelihood estimator $\hat{\alpha}_{ml}$. They found that $\hat{\alpha}_{reg}$ showed noticeable bias, while $\hat{\alpha}_{ml}$ was practically unbiased. Vermeulen \cite{vermeulen14} made Monte-Carlo experiments under the assumption of a non-response rate that was correlated with wealth according to \cite{kennickell93}. This lead to considerable bias, which was slightly stronger for $\hat{\alpha}_{ml}$ than for $\hat{\alpha}_{reg}$.

Another problem with ansatz (\ref{eq:pareto_ccdf}) is that it is not applicable in the presence of missing data, e.g.~for the combined HFCS and Manager Magazin data, which have a considerable gap between the highest HFCS and lowest Manager Magazin wealth value. Eq.~(\ref{eq:pareto_ccdf}) is based on the complementary empirical distribution function, and thus requires knowledge of the weight of the missing values, which cannot be estimated without knowing $\alpha$ beforehand. The same problem affects ansatz (\ref{eq:vanderwijk_est}) and the resulting estimator (\ref{eq:alpha_wijk}). The maximum likelihood estimator, however, does not suffer from this shortcoming, because the term $N(w_{\mbox{\scriptsize\it min}})$ in (\ref{eq:alpha_ml}) is not used as an estimator for the distribution function, but simply represents the total number of measured values.

\begin{figure}[t]
\centering
\includegraphics[width=0.8\columnwidth]{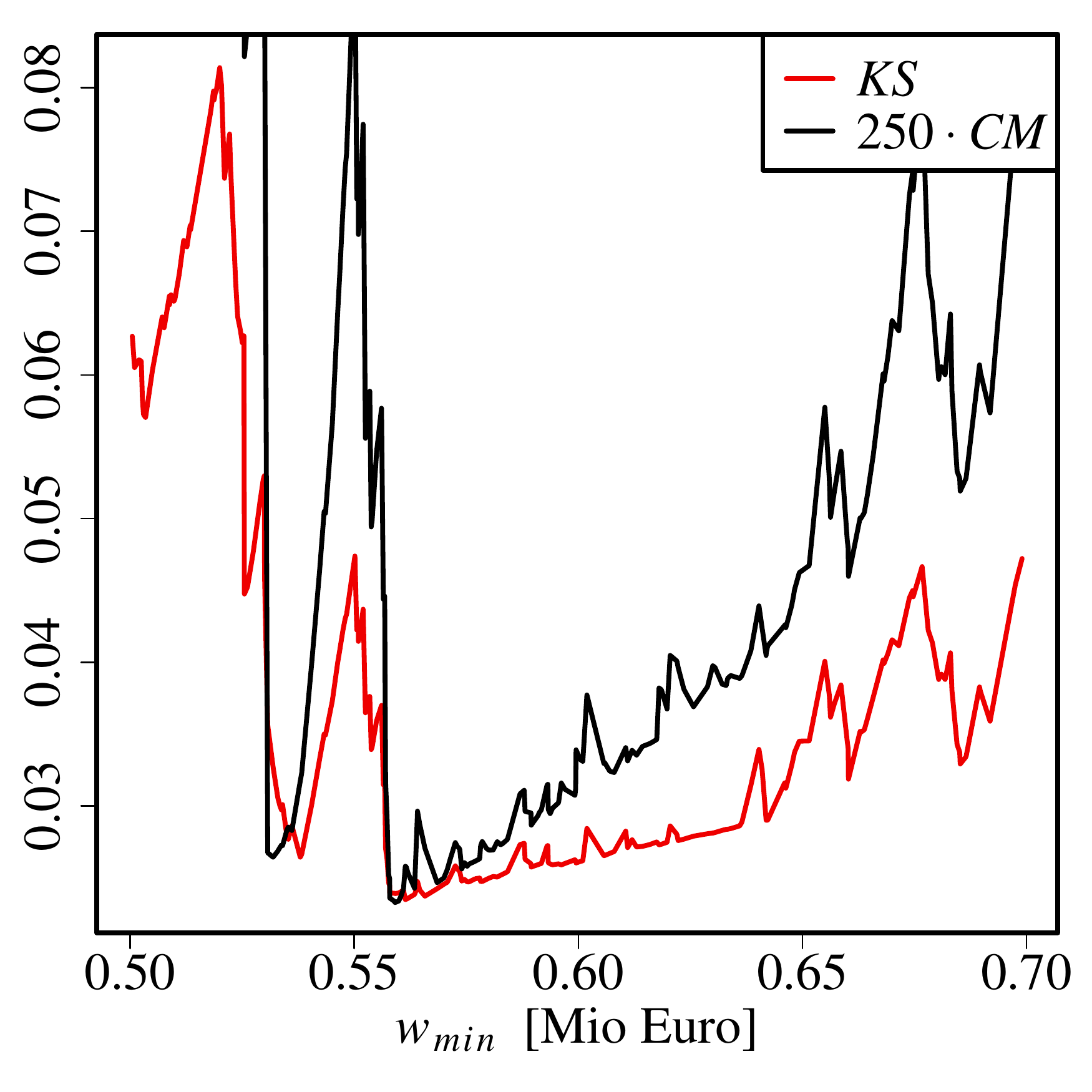}
\caption{\label{fig:ks_cm}Goodness-of-fit criteria after Kolmogorov-Smirnov ({\it KS}) and Cramer-van Mieses ({\it CM}) as a function of $w_{\mbox{\scriptsize\it min}}$ for the averaged HFCS data for Germany and $\hat{\alpha}_{ml}$. The {\it CM}-values have been multiplied by 250 to make the curves comparable.}
\end{figure}

Whichever of the estimators for $\alpha$ is used, in any case a choice for $w_{\mbox{\scriptsize\it min}}$ is necessary. Bach et al.~\cite{bach15} based their choice on a visual inspection of Fig.~\ref{fig:wijk} and set $w_{\mbox{\scriptsize\it min}}$ somewhere at the beginning of the region where the curve is roughly constant. This approach, however, does not yield a well-defined algorithm for the exact choice of $w_{\mbox{\scriptsize\it min}}$, and an optimality criterion based on the goodness-of-fit is preferable. Goodness-of-fit is typically measured with the distance between the empirical cumulative distribution function $F_{\mbox{\scriptsize\it emp}}(w)$ and the fitted cumulative distribution function $F_{\mbox{\scriptsize\it fit}}(w)$. Clauset et al.~\cite{clauset09} recommended the Kolmogorov-Smirnov criterion, which is the maximum distance between the two distribution functions (see Fig.~\ref{fig:ks}):
\begin{figure}[t]
\centering
\includegraphics[width=0.8\columnwidth]{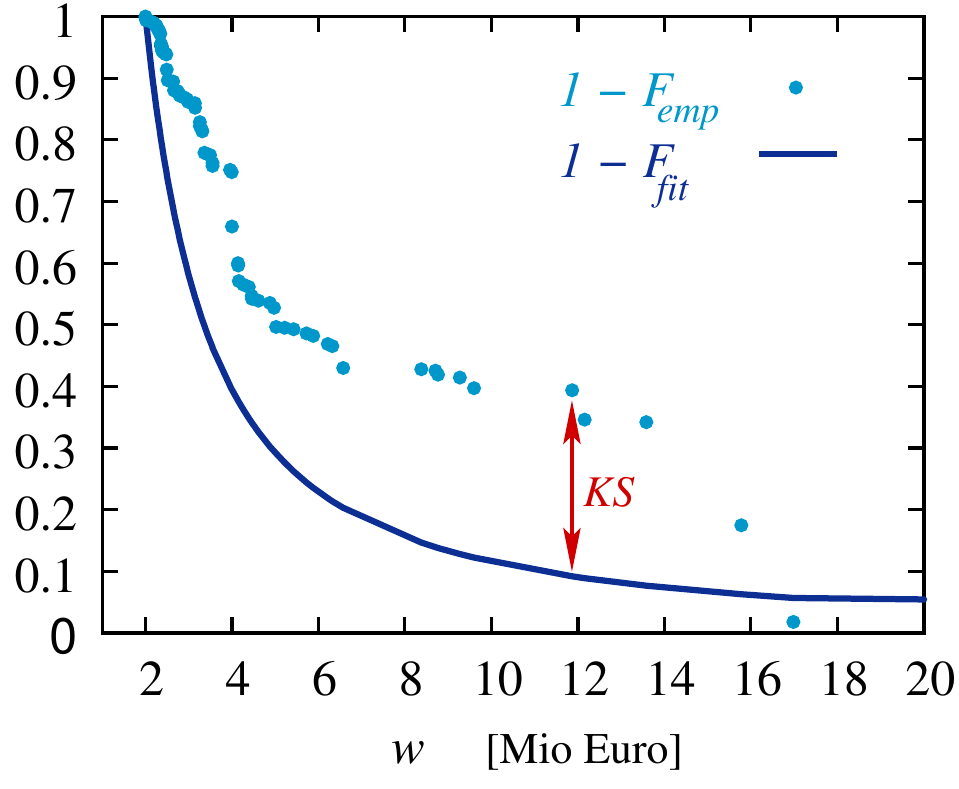}
\caption{\label{fig:ks}Goodness-of-fit criterion after Kolmogorov-Smirnov (KS).}
\end{figure}
\begin{eqnarray}
\mbox{\it KS} & = & \max_{w\geq w_{\mbox{\scriptsize\it min}}} \left| F_{\mbox{\scriptsize\it fit}}(w) - F_{\mbox{\scriptsize\it emp}}(w) \right| \\
& = & \max_{w_i\geq w_{\mbox{\scriptsize\it min}}} \left| \left(\frac{w_i}{w_{\mbox{\scriptsize\it min}}}\right)^{-\alpha} - \frac{N(w_i)}{N(w_{\mbox{\scriptsize\it min}})} \right| \nonumber
\end{eqnarray}
while Eckerstorfer et al.~\cite{eckerstorfer15} used the Cramer-van Mieses criterion, which is based on the area between the distribution functions:
\begin{equation}
\label{eq:cm_def}
\mbox{\it CM} = \int_{w_{\mbox{\scriptsize\it min}}}^\infty \Big(F_{\mbox{\scriptsize\it fit}}(w) - F_{\mbox{\scriptsize\it emp}}(w)\Big)^2 f_{\mbox{\scriptsize\it fit}}(w)\,dw
\end{equation}
When the integral is numerically evaluated with the trapezoidal rule, the criterion becomes
\begin{equation}
\mbox{\it CM} = \sum_{w_i\geq w_{\mbox{\scriptsize\it min}}} (w_{i+1}-w_i)\,\frac{g(w_i)+g(w_{i+1})}{2}
\end{equation}
where $g(w)$ is the integrand in (\ref{eq:cm_def}), i.e.
\begin{displaymath}
g(w_i) = \left( \left(\frac{w_i}{w_{\mbox{\scriptsize\it min}}}\right)^{-\alpha} \! - \frac{N(w_i)}{N(w_{\mbox{\scriptsize\it min}})}\right)^2 w_i^{-(\alpha+1)}
\end{displaymath}
As can be seen in Fig.~\ref{fig:ks_cm}, both criteria have typically the same qualitative dependency on $w_{\mbox{\scriptsize\it min}}$ with a minimum at almost the same position. This means that both criteria yield very similar optimal choices for $w_{\mbox{\scriptsize\it min}}$.

\subsection{Normalization constant $C$}
\label{sec:parameter:c}
The normalization constant $C$ in the Pareto distribution $f(w) = C\cdot w^{-(\alpha+1)}$ for $w>w_0$ must be chosen such that the density combined from HFCS-histogram and Pareto-fit is normalized to one, i.e.~$\int_{-\infty}^\infty f(w)\,dw=1$. Bach et al.~\cite{bach15} have achieved this by setting the weight of the Pareto tail equal to the HFCS weight of all households with wealth $w>w_0$:
\begin{align}
&\frac{1}{N}\sum_{w_i > w_0} n(w_i) = \int_{w_0}^\infty C\,w^{-\alpha-1} dw  \nonumber\\
\Rightarrow\quad & C  = \alpha w_0^\alpha \frac{1}{N}\sum_{w_i > w_0} n(w_i) \label{eq:c_bach}
\end{align}
As the data frequency in the region around $w_0$ is typically quite low, this choice has the effect that $C$ varies in a discontinuous way as $w_0$ moves across a data point. Eckerstorfer et al.~\cite{eckerstorfer15} used the same method, but with $w_{\mbox{\scriptsize\it min}}$ (see section \ref{sec:parameter:alpha}) instead of $w_0$:
\begin{equation}
C'  = \alpha w_{\mbox{\scriptsize\it min}}^\alpha \frac{1}{N}\sum_{w_i > w_{\mbox{\scriptsize\it min}}} n(w_i)
\label{eq:c_eckerstorfer}
\end{equation}
The choice (\ref{eq:c_eckerstorfer}) for the normalization constant has the side effect that the estimated density function is no longer normalized to one, but the area under the density function depends on the choice for $w_0$. To make sure that the total number $N$ of all households remained unchanged, Eckerstorfer et al.~rescaled all HFCS weights for wealth values below $w_{\mbox{\scriptsize\it min}}$ to
\begin{equation}
\label{eq:n_eckerstorfer}
n'(w_i) = \beta'\cdot n(w_i) \quad\mbox{ for } w_i\leq w_{\mbox{\scriptsize\it min}}
\end{equation}
where
\begin{equation}
\label{eq:beta_eckerstorfer}
\beta' = \frac{N-\hspace{-1em}\sum\limits_{w_{\mbox{\scriptsize\it min}} < w_i \leq w_0}\hspace{-1.75em} n(w_i) - \left(\frac{w_{\mbox{\scriptsize\it min}}}{w_o}\right)^\alpha \hspace{-0.75em}\sum\limits_{w_i>w_{\mbox{\scriptsize\it min}}}\hspace{-1em} n(w_i)}{\sum\limits_{w_i\leq w_{\mbox{\scriptsize\it min}}}\hspace{-1em} n(w_i)}
\end{equation}
A new way to obtain yet another estimate $C''$ for the normalization constant consists in a utilization of the rich list, which gives the exact number of households $\tilde{N}_{w_1}$ above some wealth threshold $w_1$:
\begin{equation}
\label{eq:c_dalitz_pre_a}
\frac{\tilde{N}(w_1)}{N'} = C'' \int_{w_1}^\infty w^{-\alpha -1} dw
\end{equation}
where $\tilde{N}(w_1)$ is the number of households in the rich list with wealth greater or equal $w_1$. $N'$ is the total number of households, which is increased by (\ref{eq:c_dalitz_pre_a}) to
\begin{equation}
\label{eq:c_dalitz_pre_b}
N' = \sum_{w_i\leq w_0} n(w_i) + N' C'' \int_{w_0}^\infty w^{-\alpha -1} dw
\end{equation}
Evaluating the integrals in (\ref{eq:c_dalitz_pre_a}) and (\ref{eq:c_dalitz_pre_b}) and solving for $C''$ and $N'$ yields
\begin{eqnarray}
\label{eq:c_dalitz_a}
N' & = & \sum_{w_i\leq w_0} n(w_i) + \tilde{N}(w_1) \left(\frac{w_1}{w_0}\right)^\alpha \\
\label{eq:c_dalitz}
C'' & = & \frac{\tilde{N}(w_1)}{N'} \alpha w_1^\alpha
\end{eqnarray}
The threshold $w_1$ should be chosen as low as possible, yet still in the region where van der Wijk's law holds for the rich list data. For the Manager Magazin data, e.g., it can be concluded from Fig. \ref{fig:wijk} that $w_1 \approx 500$ Mio \euro{} is a reasonable threshold for this rich list. As wealth values in this list are only rough estimates and the gap to the next wealth value is 50 Mio \euro{}, the most conservative choice is to use $\tilde{N}(w_1)$ with $w_1 = 500$ Mio \euro{} and $w_1^\alpha$ with $w_1 = 450$ Mio \euro.

\subsection{Transition threshold $w_0$}
\label{sec:parameter:w_0}
Little attention has been paid in the literature to the choice of the transition threshold $w_0$. From the discussion in section \ref{sec:parameter:alpha}, it would seem natural to set it equal to $w_{\mbox{\scriptsize\it min}}$, i.e., to the threshold that yields the best fit according to a goodness-of-fit criterion. Surprisingly, both Bach et al.~and Eckerstorfer et al.~set it much higher: $w_{\mbox{\scriptsize\it min}}=0.5$ Mio \euro{} \& $w_0=3$ Mio \euro{} (Bach et al.~for Germany \cite{bach15}) and $w_{\mbox{\scriptsize\it min}}\approx 0.3$ Mio \euro{} \& $w_0=4$ Mio \euro{} (Eckerstorfer et al.~for Austria \cite{eckerstorfer15}).

Eckerstorfer et al.~argued: ``We choose this \euro{4} million cut-off point because the frequency of observations starts to markedly decline beyond this level of net wealth.'' They also observed that changing $w_0$ to 3 Mio \euro{} or 5 Mio \euro{} only had a minor impact on the final results. This does not explain, however, why $w_0$ should be limited to this range. As can be seen from Table \ref{tbl:w_0_effect}, there actually are noticeable differences for the resulting top percent share as $w_0$ varies. It is therefore desirable to have a well-defined criterion for choosing $w_0$.

\begin{table}[t]
\centering
\begin{small}
\begin{tabular}{c|cc}
$w_0$ [Mio \euro] & $s_{0.01}$ with $C$ & $s_{0.01}$ with $C'$ \\\hline
$w_{\mbox{\scriptsize\it min}}$ & 0.292  & 0.292 \\
1.0 & 0.290 & 0.292 \\
3.0 & 0.303 & 0.283 \\
5.0 & 0.266 & 0.301 \\
7.0 & 0.279 & 0.292 \\
9.0 & 0.303 & 0.281 \\
11.0 & 0.322 & 0.274
\end{tabular}
\end{small}
\caption{\label{tbl:w_0_effect}Impact of the choice of $w_0$ on the resulting top percent wealth share $s_{0.01}$ for the averaged HFCS data for Germany with $w_{\mbox{\scriptsize\it min}}=561400$ \euro{} and $\alpha=1.4735$. The weight normalization has been made with $C$ given by Eq.~(\ref{eq:c_bach}), and with $C'$ given by Eqs.~(\ref{eq:c_eckerstorfer}) to (\ref{eq:beta_eckerstorfer}).}
\end{table}

A natural way to restrict the choice of $w_0$ is by imposing the demand of continuity upon the estimated density function $\hat{f}(w)$. As can be seen from Fig.~\ref{fig:densityapprox}, this is not a reasonable assumption for the histogram estimator from the HFCS data, which is discontinuous itself. It is reasonable, however, for the kernel density estimator $\hat{f}_{\mbox{\scriptsize\it kern}}(w)$ given by Eq.~(\ref{eq:f_kernel}). This leads to the ansatz
\begin{equation}
\label{eq:w0_dalitz}
\hat{f}_{\mbox{\scriptsize\it kern}}(w_0) - C\cdot w_0^{-(\alpha+1)} = 0
\end{equation}
which is to be solved for $w_0$ numerically. As Eq.~(\ref{eq:w0_dalitz}) generally has more than one solution $w_0$, the smallest zero greater than $w_{\mbox{\scriptsize\it min}}$ should be chosen. For the averaged HFCS data for Germany and the normalization with (\ref{eq:c_bach}) or (\ref{eq:c_eckerstorfer}), this leads to $w_0 \approx 830\ 000$ \euro{} when the kernel density bandwidth is automatically selected from all data points greater than $w_{\mbox{\scriptsize\it min}}$ with the method by Sheather \& Jones\footnote{This is implemented as method ``SJ'' in the {\em R} function {\em density}} \cite{sheather91}, which yields a bandwidth of 55 551 \euro{}.

\begin{table*}[b]
\centering
\begin{small}
\begin{tabular}{cc|c|ccccc}
{\em cutoff} & {\em weighted} & {\em type} & {\em mean} & $\sqrt{s^2}$ & MSE & KS & CM \\\hline
 & & $\hat{\alpha}_{ml}$ & 1.4997 & 0.0215 & 0.000462 & 0.010246 & 0.000018 \\
 no & no & $\hat{\alpha}_{reg}$ & 1.4956 & 0.0238 & 0.000587 & 0.011865 & 0.000030 \\
 & & $\hat{\alpha}_{reg'}$ & 1.4924 & 0.0294 & 0.000924 & 0.013691 & 0.000052 \\
 & & $\hat{\alpha}_{\mbox{\scriptsize\it wijk}}$ & 1.5108 & 0.0552 & 0.003159 & 0.017555 & 0.000128 \\\hline
 & &  $\hat{\alpha}_{ml}$ & 1.5061 & 0.0213 & 0.000491 & 0.010484 & 0.000020 \\
 yes & no & $\hat{\alpha}_{reg}$ & 1.5106 & 0.0228 & 0.000632 & 0.012041 & 0.000031 \\
 & & $\hat{\alpha}_{reg'}$ & 1.5171 & 0.0279 & 0.001073 & 0.013736 & 0.000050 \\
 & & $\hat{\alpha}_{\mbox{\scriptsize\it wijk}}$ & 1.5707 & 0.0237 & 0.005551 & 0.022747 & 0.000190 \\\hline
 & & $\hat{\alpha}_{ml}$ & 1.5005 & 0.0003 & 0.000000 & 0.001000 & 0.000000 \\
 no & yes & $\hat{\alpha}_{reg}$ & 1.4971 & 0.0006 & 0.000009 & 0.000765 & 0.000000 \\
 & & $\hat{\alpha}_{reg'}$ & 1.4954 & 0.0011 & 0.000022 & 0.001124 & 0.000001 \\
 & & $\hat{\alpha}_{\mbox{\scriptsize\it wijk}}$ & 1.5309 & 0.0087 & 0.001031 & 0.008186 & 0.000035 \\\hline
 & & $\hat{\alpha}_{ml}$ & 1.5063 & 0.0001 & 0.000040 & 0.001975 & 0.000001 \\
 yes & yes & $\hat{\alpha}_{reg}$ & 1.5109 & 0.0025 & 0.000125 & 0.003070 & 0.000004 \\
 & & $\hat{\alpha}_{reg'}$ & 1.5173 & 0.0040 & 0.000315 & 0.004595 & 0.000009 \\
 & & $\hat{\alpha}_{\mbox{\scriptsize\it wijk}}$ & 1.5708 & 0.0018 & 0.005017 & 0.017273 & 0.000154 \\
\end{tabular}
\end{small}
\caption{\label{tbl:simulations}Mean, standard deviation $\sqrt{s^2}$, MSE, and goodness-of-fit criteria KS and CM of the three estimators for data randomly drawn from a Pareto distribution with true value $\alpha=1.5$. $\hat{\alpha}_{reg}$ is the proper linear regression estimator given by Eq.~(\ref{eq:alpha_reg}), and $\hat{\alpha}_{reg'}$ is a linear regression estimator with an additional intercept term.}
\end{table*}

\section{Results}
\label{sec:results}
Section \ref{sec:parameter:alpha} gives different possible estimators for the power $\alpha$, and the Monte-Carlo study \cite{clauset09} favors the maximum-likelihood estimator $\hat{\alpha}_{ml}$ given by (\ref{eq:alpha_ml}) over the regression estimator $\hat{\alpha}_{reg}$ given by (\ref{eq:alpha_reg}), while \cite{vermeulen14} favors the regression estimator. The estimator based on van der Wijk's law $\hat{\alpha}_{\mbox{\scriptsize\it wijk}}$ given by (\ref{eq:alpha_wijk}) has not yet been studied in the literature.

Therefore, bias and mean-squared-error of these three estimators was first compared in Monte-Carlo experiments. The best performing estimator in these experiments, which was $\hat{\alpha}_{ml}$, was then used to estimate both the power $\alpha$ and the wealth share of the richest percent.

\subsection{Estimator comparison by simulations}
\label{sec:results:simulation}
To compare the three estimators, 5000 sample values were drawn from a Pareto distribution with the function {\em rpareto} from the {\em R} package {\em actuar} \cite{actuar08}. The parameter $\alpha$ was set to 1.5 because this had turned out to be a cross-country constant in different studies according to Gabaix \cite{gabaix09}, and $w_{\mbox{\scriptsize\it min}}$ was set to 0.5 Mio \euro. This was repeated 1000 times to estimate bias and mean squared error (MSE) of the three estimators. Two additional variants of the experiments were implemented:
\begin{itemize}
\item to simulate the cutoff in the HFCS data, values greater than 75 Mio \euro{} were rejected
\item to simulate the HFCS weighting process, weights were added that represent the exact area under the Pareto density in a cell around each sample value
\end{itemize}
As both variants could be applied or not, this lead to four different simulations. The results are shown in Table \ref{tbl:simulations}: the maximum-likelihood estimator $\hat{\alpha}_{ml}$ showed the smallest MSE in all cases, and the estimator based on van der Wijk's law the largest. According to these results, the maximum-likelihood estimator is the most preferable.

It is interesting to note that adding an intercept term in the linear regression did not improve the least squares fit, but lead both to a larger MSE and and to poorer goodness-of-fit values compared to a linear regression without an intercept term (see the values for $\hat{\alpha}_{reg'}$ in Table \ref{tbl:simulations}). This seems surprising at first sight because one would expect that an additional fit parameter improves the fit. This does not hold in this case, however,  because the model with the additional parameter no longer represents the underlying distribution.

\begin{table*}[t]
\centering
\begin{small}
\begin{tabular}{c|ccc|ccc}
 & \multicolumn{3}{|c|}{\em averaged HFCS} & \multicolumn{3}{|c}{\em Manager Magazin} \\
{\em estimator} & {\em goodness-of-fit} & $\alpha$ & $w_{\mbox{\scriptsize\it min}}$ & {\em goodness-of-fit} & $\alpha$ & $w_{\mbox{\scriptsize\it min}}$ \\\hline

 & {\em KS} = 0.023479 & 1.4735 & 561400 \euro & {\em KS} = 0.057548 & 1.5168 & 500 Mio \euro \\ 
\raisebox{1.5ex}[-1.5ex]{$\hat{\alpha}_{ml}$} & {\em CM} = 0.000093 & 1.4877 & 559100 \euro & {\em CM} = 0.000764 & 1.5777 & 550 Mio \euro  \\\hline

 & {\em KS} = 0.025494 & 1.5179 & 538292 \euro & {\em KS} = 0.077164 & 1.4377 & 500 Mio \euro \\ 
\raisebox{1.5ex}[-1.5ex]{$\hat{\alpha}_{reg}$} & {\em CM} = 0.000111 & 1.5091 & 557932 \euro & {\em CM} = 0.000679 & 1.4377 & 500 Mio \euro  \\\hline

& {\em KS} = 0.028566 & 1.5315 & 538292 \euro & {\em KS} = 0.078456 & 1.4326 & 500 Mio \euro \\
\raisebox{1.5ex}[-1.5ex]{$\hat{\alpha}_{reg'}$} & {\em CM} = 0.000133 &  1.5313 & 535300 \euro & {\em CM} = 0.000687 & 1.4326 & 500 Mio \euro \\\hline

& {\em KS} = 0.040020 & 1.6032 & 546800 \euro & {\em KS} = 0.066083 & 1.6255 & 550 Mio \euro  \\
\raisebox{1.5ex}[-1.5ex]{$\hat{\alpha}_{\mbox{\scriptsize\it wijk}}$} & {\em CM} = 0.000282 & 1.5955 & 540500 \euro & {\em CM} = 0.000893 & 1.6255 & 550 Mio \euro \\
\end{tabular}
\end{small}
\caption{\label{tbl:estimators}Comparison of the results of the different estimators different goodness-of-fit criteria for estimating $\alpha$ and $w_{\mbox{\scriptsize\it min}}$ on the averaged HFCS data for Germany and the Manager Magazin rich list. $\hat{\alpha}_{reg}$ is the proper linear regression estimator given by Eq.~(\ref{eq:alpha_reg}), and $\hat{\alpha}_{reg'}$ is a linear regression estimator with an additional intercept term.}
\end{table*}

\subsection{Estimation of $\alpha$}
\label{sec:results:alpha}
Table \ref{tbl:estimators} shows the results of the different estimators on the averaged HFCS data for Germany. The values for $w_{\mbox{\scriptsize\it min}}$ have been determined by selecting the best fit (highest {\em KS}, or {\em CM}, respectively) from all wealth values $w_i$ in the data. The ranking of the different estimators is the same as in the simulations in the preceding section: the maximum-likelihood estimator shows the best goodness-of-fit with respect to both criteria, and the estimator based on van der Wijk's law shows the poorest goodness-of-fit. In agreement with Fig.~\ref{fig:ks_cm}, both goodness-of-fit criteria yield similar results for $w_{\mbox{\scriptsize\it min}}$. It should be noted that the value $\alpha\approx 1.53$ reported by Bach et al.~in \cite{bach15} is the value obtained by linear regression {\em with an intercept term}.

Table \ref{tbl:estimators} also gives the results for the Manager Magazin rich list for Germany. It is interesting to note that $\hat{\alpha}_{ml}$ is higher than for the HFCS data, while $\hat{\alpha}_{reg}$ is lower, albeit not as low as reported by Bach et al.~in \cite{bach15}. This discrepancy stems from the way in which Bach et al.~used the rich list: they did not aggregate the household data by wealth, such that each value only had weight one, with the same wealth listed as often as the number of households. This has no effect on $\hat{\alpha}_{ml}$, but introduced systematic bias into $\hat{\alpha}_{reg}$. The results reported for $\hat{\alpha}_{reg}$ by Bach et al.~were therefore systematically too small when the Manager Magazin data was included in their calculation.

The estimators for $\alpha$ obtained from the rich list are less reliable than the estimators obtained from the HFCS data for two reasons:
\begin{itemize}
\item The rich list data with $w_i\geq w_{\mbox{\scriptsize\it min}}$ only represents about 200 households, while the HFCS data with $w_i \geq w_{\mbox{\scriptsize\it min}}$ represents about 2.5 Mio households. This does not only make the fitting process less robust, but also means that missing data in the rich list have a stronger effect.
\item The wealth values in the rich list are only rough estimates and there is a considerable gap between $w_i=w_{\mbox{\scriptsize\it min}}$ and $w_{i-1}$. This makes the exact placement of the threshold $w_{\mbox{\scriptsize\it min}}$ and thus also the estimator for $\alpha$ less accurate than for the tightly lying HFCS data. The choice $w_{\mbox{\scriptsize\it min}}=475$ Mio \euro{} instead of 500 Mio \euro{}, e.g., which includes exactly the same number of households because it is the mid point between 500 Mio \euro{} and the next reported wealth, leads to $\hat{\alpha}_{ml}=1.407288$.
\end{itemize}
It is therefore advisable to use the rich list data only in combination with the HFCS data, thereby leading to a refinement of the estimator obtained from the HFCS data. For reasons explained in section \ref{sec:parameter:alpha}, only the maximum-likelihood estimator $\hat{\alpha}_{ml}$ can be used for combined data.

\begin{table}[b!]
\centering
\begin{small}
\begin{tabular}{c|c@{\hskip 0.5ex}c|cc}
{\em HFCS} & \multicolumn{2}{|c|}{$w_{\mbox{\scriptsize\it min}}$} & {$\hat{\alpha}_{ml}$} & {$\hat{\alpha}_{ml}$} \\
{\em variant} & \multicolumn{2}{|c|}{\em criterion} & HFCS & {\em combined} \\\hline
 & {\em KS} & $\to$ 561 400 \euro & 1.4735 & 1.4723 \\
\raisebox{1.5ex}[-1.5ex]{avg} & {\em CM} & $\to$ 559 100 \euro & 1.4877 & 1.4865 \\\hline
 & {\em KS} & $\to$ 561 500 \euro & 1.4664 & 1.4652 \\
\raisebox{1.5ex}[-1.5ex]{1} & {\em CM} & $\to$ 558 000 \euro  & 1.4810 & 1.4798 \\\hline
 & {\em KS} & $\to$ 564 030 \euro & 1.4802 & 1.4790 \\
\raisebox{1.5ex}[-1.5ex]{2} & {\em CM} & $\to$ 563 430 \euro & 1.4827 & 1.4814 \\\hline
 & {\em KS} & $\to$ 597 000 \euro & 1.4680 & 1.4666 \\
\raisebox{1.5ex}[-1.5ex]{3} & {\em CM} & $\to$ 595 800 \euro & 1.4691 & 1.4677 \\\hline
 & {\em KS} & $\to$ 619 800 \euro & 1.4499 & 1.4485 \\
\raisebox{1.5ex}[-1.5ex]{4} & {\em CM} & $\to$ 618 000 \euro & 1.4604 & 1.4590 \\\hline
 & {\em KS} & $\to$ 561 500 \euro & 1.4454 & 1.4443 \\
\raisebox{1.5ex}[-1.5ex]{5} & {\em CM} & $\to$ 561 500 \euro & 1.4622 & 1.4611
\end{tabular}
\end{small}
\caption{\label{tbl:alpha_combined}Maximum-likelihood estimator for $\alpha$ obtained from the combined HFCS ($w_i \geq w_{\mbox{\scriptsize\it min}}$) and Manager Magazin ($w_i\geq 500$ Mio \euro{}) data. ``HFCS variant'' means the five differently imputed datasets and the averaged imputation.}
\end{table}

The results for the combined fits are shown in Table \ref{tbl:alpha_combined}. Note that the ``average'' value is not the average of the values given below, but the maximum-likelihood estimator obtained from the averaged individual wealth values. That the resulting estimator from the averaged data with the Cramer-van Mieses goodness-of-fit criterion is greater than the estimator for each individual dataset is surprising, but can be explained by the highly non-linear relationship between the $w_i$ and $w_{\mbox{\scriptsize\it min}}$ and $\hat{\alpha}_{ml}$.

Taking the Manager Magazin rich list into account decreases $\alpha$, albeit only marginal: these changes to the fourth significant digit are smaller than the variation due to the variety of imputations, which affects the third significant digit. Bach et al.~reported in \cite{bach15} a much greater decrease of $\alpha$ when the rich list data was taken into account (from 1.535 to 1.370), but their results were based on a least squares fit on the data points $N(w_i) =\sum_{w_j\geq w_i} n(w_j)$ with the assumption that the missing data between the highest HFCS and the lowest Manager Magazin wealth had zero weight. Such an assumption leads to additional bias of the least squares fit estimator $\hat{\alpha}_{reg}$. The maximum-likelihood estimators reported in Table \ref{tbl:alpha_combined} do not suffer from this problem and lead to the conclusion that $\alpha$ lies between 1.443 and 1.481 for Germany, depending on the imputation variant.

\subsection{Estimation of one percent share}
\label{sec:results:share}
Table \ref{tbl:share} shows the estimated wealth share of the richest $p$ percent in Germany when the different ways for normalizing the Pareto distribution are applied: $C$ stands for the method by Bach et al.~given by Eq.~(\ref{eq:c_bach}), $C'$ for the method by Eckerstorfer et al.~given by Eq.~(\ref{eq:c_eckerstorfer}), and $C''$ for the new method utilizing the rich list as given by Eq.~(\ref{eq:c_dalitz}). The transition thresholds $w_0$ have been determined with the continuity condition (\ref{eq:w0_dalitz}). These have been computed for all five imputed datasets and the averaged dataset separately, but are given in Table \ref{tbl:share} only for the averaged dataset.

\begin{table}[t]
\centering
\begin{small}
\begin{tabular}{r|ccc}
& \multicolumn{3}{|c}{\em normalization method} \\
& $C$ & $C'$ & $C''$ \\\hline
\multicolumn{1}{c|}{$w_0$ {\em (avg)}} & 820 790 & 839 135 & 1 791 065 \\\hline
{\em (avg)} & 0.292 & 0.293 & 0.334 \\
$s_{0.01}$ {\em (min)} & 0.287 & 0.284 & 0.323 \\
{\em (max)} & 0.305 & 0.301 & 0.344 \\\hline
{\em (avg)} & 0.493 & 0.491 & 0.565 \\
$s_{0.05}$ {\em (min)} & 0.492 & 0.479 & 0.529 \\
{\em (max)} & 0.504 & 0.494 & 0.567 \\\hline
{\em (avg)} & 0.619 & 0.617 & 0.675 \\
$s_{0.10}$ {\em (min)} & 0.619 & 0.609 & 0.661 \\
{\em (max)} & 0.628 & 0.621 & 0.690 \\
\end{tabular}
\end{small}
\caption{\label{tbl:share}Wealth share of the richest $p\in\{0.01, 0.05, 0.10\}$ percent when the $\alpha$ and $w_{\mbox{\scriptsize min}}$ values from the rightmost column for goodness-of-fit criterion {\em KS} in Table \ref{tbl:alpha_combined} are used. {\em avg} is the result for the averaged data, {\em min/max} the minimum/maximum share over all five datasets. The normalization methods refer to section \ref{sec:parameter:c}.}
\end{table}

The results for normalizations $C$ and $C'$ are comparable, but the shares obtained with the normalization based on the rich list are 4 to 6 percentage points higher. This is not unexpected, because the normalization methods by Eckerstorfer et al.~and by Bach et al.~both remove weights above $w_{\mbox{\scriptsize min}}$ and assign it to higher wealth. The underlying assumption is that the total number of weights assigned to wealths greater than $w_{\mbox{\scriptsize min}}$ (or greater than $w_0$ in Bach's case) is estimated correctly in the HFCS data. The new method based on the rich list assumes, however, that the number of households in the rich list is correct instead, which means that the total number of households with high wealth is estimated too low in the HFCS data. The author considers the latter to be a more realistic assumption; actually this is the main reason for using the rich list at all.

\section{Conclusions}
\label{sec:conclusions}
The maximum-likelihood estimator for the power $\alpha$ of the Pareto distribution showed the best performance both in the Monte-Carlo simulations and the goodness-of-fit tests on the HFCS data, and it is therefore recommended for fitting a Pareto distribution to wealth survey data. Combining the HFCS data with data from a rich list has little impact on the maximum-likelihood estimator for $\alpha$. Neither does the resulting wealth share of the richest percent increase when rich list data is taken into account only for estimating the power $\alpha$. If the rich list is used, however, for estimating the normalization constant $C$ in the Pareto distribution, the wealth share increases about four percentage points in comparison to an estimation of this normalization constant from the HFCS data.

The new method for determining the transition threshold $w_0$ by imposing a continuity condition on the wealth distribution density provides a way for removing the arbitrariness of this parameter. It relies on a kernel density estimator for the wealth density. Theoretically, this kernel density estimator can also be utilized for computing the top percent wealth share by numeric integration. It would be interesting to investigate whether this would make the dependency of the wealth share on the choice of $w_0$ less noisy.

\section*{Acknowledgments}
The author is grateful to the ECB for providing the HFCS survey data, and to Andreas Thiemann from the DIW for providing the extended Manager Magazin rich list and for explaining some aspects of the DIW study \cite{bach15}.

\bibliographystyle{ieeetr}
\bibliography{wealth-distribution}

\end{document}